\documentstyle[preprint,aps,pre]{revtex} 
\begin{document}                          
\draft                                    


\title{Island diffusion on metal fcc(100) surfaces}

\author{J. Heinonen$^{1,\dagger}$,
        I. Koponen$^1$, 
        J. Merikoski$^2$ and
        T. Ala-Nissila$^{1,3}$}
 
\address{
$^1$ Helsinki Institute of Physics, 
     P.O. Box 9, FIN--00014 University of Helsinki, Helsinki, Finland\\
$^2$ Department of Physics, University of Jyv\"askyl\"a,
     P.O. Box 35, FIN--40351 Jyv\"askyl\"a, Finland\\
$^3$ Laboratory of Physics, Helsinki University of Technology,
     P.O. Box 1100, FIN--02150 HUT, Espoo, Finland, and Department of
     Physics, Brown University, Providence R.I. 02912--1843}

\date{October 9, 1998}

\maketitle

\begin{abstract}

We present Monte Carlo simulations for the size and temperature
dependence of the diffusion coefficient of adatom islands
on the Cu(100) surface.
We show that the scaling exponent for the size dependence is not a constant
but a decreasing function of the island size and approaches unity for very
large islands.
This is due to a crossover from periphery dominated
mass transport to a regime where vacancies diffuse inside the island.
The effective scaling exponents are in good agreement with theory and
experiments.

\end{abstract}

\bigskip

Theoretical studies on island diffusion over the past two
decades have lead to expectations that even large islands may 
have substantial mobilities \cite{Seminal,Vot86}. 
A seminal study of diffusion of large islands on metallic surfaces was
done by Voter \cite{Vot86}, where he was able to show 
that the diffusion coefficient $D$ of islands with more
than $s \approx 10$ atoms followed a simple
scaling law with a constant scaling exponent $\alpha$

\begin{equation} 
\label{scaling}
D \propto e^{-\beta E_{L}}  \; s^{-\alpha},
\end{equation}

\noindent
where $\beta=1/(k_{B}T)$ and $E_L$ is an effective energy barrier
for island diffusion.

Since then similar scaling law for large islands, 
with the scaling exponent $\alpha$ now depending on
the diffusion mechanism, has been found in several simulation studies 
\cite{Kan90,Sho95,Bog98}. However, the experimental 
confirmation of the early theoretical predictions had to wait for 
the development of advanced scanning tunneling microscope (STM)
techniques. Only recently the experiments 
have unequivocally confirmed that on metal
surfaces even large islands of sizes up to 1000 atoms undergo 
diffusion and that the diffusion coefficient obeys Eq. (\ref{scaling})
with $\alpha$ indeed depending on the diffusion mechanism \cite{Wen94,Pai97}. 
Although the experiments and simulations have given strong support to
the scaling law in Eq. (\ref{scaling}), at least in a restricted region
of sizes, the exact role of the various microscopic mechanisms
in determining the value of $\alpha$ is still an open question.

On the theoretical side, Khare {\it et al.} \cite{Kha95,Kha96}
have explained island diffusion in terms  of the shape fluctuations
of the outer boundary, which makes it possible to relate
the macroscopic motion of islands
to the atomistic processes occurring on the boundary. 
The three basic mechanisms considered are
particle diffusion along the periphery (PD), terrace diffusion (TD)
where a particle can detach from and attach to the edge,
and evaporation and condensation
limited diffusion mechanism (EC).
The effective exponent
$\alpha(R) \equiv -{{\partial \ln (D)}/{\partial \ln (R)}}$
can be expressed as
\cite{Kha95,Kha96}

\begin{eqnarray}
\label{alpha}
2\alpha  & = & 2 +{{1}\over{1+(R/R_{st})(R_{su}/R_{st})}}
\nonumber\\ 
\nonumber\\ 
 &&-{{2+(R/R_{st})(R_{su}/R_{st})}
\over{1+(R/R_{st})(R_{su}/R_{st})+(R/R_{st})^2}},
\end{eqnarray}

\noindent
where $R=\sqrt{s/ \pi}$.
The parameters $R_{st}$ and $R_{su}$ are related to
periphery and terrace diffusion coefficients, respectively.
Allowing only one of the
mass transport mechanisms EC, TD or PD
at a time for large enough islands,
the exponents 1/2, 1, 3/2 are obtained, respectively
(see Fig 3. in Ref. \cite{Kha96}).
When both the TD and the PD mechanisms are present,
clear dependence of
$\alpha$ on $s$ should be observed, and finally 
one should always find $\alpha$ = 1 for $s \rightarrow \infty$.
However, the crossover regime towards this limit
occupies rather narrow region in the
parameter space and it has been assumed to be experimentally
unaccessible \cite{Kha96,Kha98}.

In contrast,
most simulations of island diffusion on metallic 
fcc surfaces indicate values $1.75 < \alpha < 2.1$ 
\cite{Vot86,Sho95,Bit95} that cannot be obtained from
the theory of Khare {\it et al.} \cite{Kha95,Kha96}.
However, their approach is strongly supported by
the recent experiments of Pai {\it et al.} \cite{Pai97},
whose careful STM measurements on 
the diffusion of Cu and Ag islands on Cu(100) and Ag(100)
surfaces yielded
$\alpha \approx 1.25$ and $\alpha \approx 1.14$, respectively,
at room temperature.
According to their explanation, these values of $\alpha$ are due to
the lack of the TD mechanism with
$R_{su}=0$, and $0.1 < R/R_{st} < 10$ in Eq. (\ref{alpha}).
The parameter $R_{st}$ was interpreted as the average separation
between adjacent kinks. However, the STM measurements were not
able to directly confirm the nature of microscopic diffusion
mechanisms for the islands.

In this report we will show through extensive simulations of
a realistic model of Cu islands on the Cu(100) surface
that these open questions can be resolved. First, 
our simulations show that there exist a long crossover
towards $\alpha=1$ for this system.
This indicates that large
{\it effective} values of $\alpha$
may be obtained if only relatively small 
island sizes are considered. This may explain
some of 
the large values reported in the literature in cases where there
are no unusual diffusion mechanisms present \cite{Kan90}.
Second, we show that this crossover is actually due to PD dominated
diffusion changing over to TD dominated case, where the microscopic
mechanism for the TD process comes from {\it vacancy diffusion} within 
large islands. In this way, the values of $\alpha$
obtained in Ref. \cite{Pai97} can be explained with the existence of
{\it both} PD and TD mechanisms for Cu 
islands, with vacancy diffusion now accounting for the latter. 
We also discuss the origin of
persistent oscillations in $D$ for small island sizes,
and vacancy island diffusion on the Cu(100) surface.

The model system we consider here is based on
kinetic Monte Carlo simulations of Cu adatoms on the
Cu(100) surface, with energetics
obtained from molecular dynamics simulations with the effective
medium theory (EMT) potential \cite{Mer97}. As discussed
in detail in Refs. \cite{Mer97}, the EMT barriers are in good
agreement with available experimental data for this case. 
The hopping rate $\nu$ of an atom to a vacant nearest neighbor (NN)
site can be well approximated by \cite{Mer97,Mer98}

\begin{equation}
\label{energetics}
\nu=\nu_0 e^{-\beta[E_S + \min(0,\Delta_{NN}) E_B] },
\end{equation}

\noindent
where the attempt frequency $\nu_0=3.06 \times 10^{12}$ s$^{-1}$ and
the barrier for the jump of a single adatom $E_S=0.399$ eV.
When there is at least one atom diagonally next to the saddle point
the barrier $E_{S}=0.258$ eV.
The change in the bond number $-3 \le \Delta_{NN} \le 3$
is the number of NN bonds in the final site subtracted by the number
of NN bonds in the initial site. The bond energy $E_B=-0.260$ eV.
We note that within the EMT, 
barriers on the Ag(100) and Ni(100) surfaces are
very similar to the barriers on Cu(100) up to a  
scaling factor \cite{Mer97}. We therefore expect that the features
observed here may describe island diffusion on 
some other fcc(100) metal surfaces, too. 

In this work we prevent detachment of adatoms 
from the island, however, an 
adatom can still go around the corner so that the PD
mechanism is operational \cite{Vot86,Bog98}.
It thus follows that $E_{S}=0.258$ eV for all the allowed jumps.
Therefore, the energetics in Eq. (\ref{energetics}) for the adatom islands
is equivalent to the ferromagnetic Ising model with Metropolis transition
rates and Kawasaki dynamics.

We create the initial island of
$s$ particles by adding
atoms one by one to the nearest and the next nearest neighbor
sites with the probability $\propto e^{-\beta z E_B}$,
where $0 \le z \le 4$ is the number of nearest neighbors. It is important to
start the simulation with a well thermalized island configuration since
the relaxation times for larger islands can become very long.
After thermalization,
we compute the tracer diffusion coefficient of the island
defined through
$D = \lim_{\displaystyle t \to \infty}  
{{1}\over{4}} d \langle r^2 \rangle / dt $
where $\langle r^2 \rangle$ is the mean square displacement of the
island \cite{Vot86}. An efficient way of computing $D$ is given
in Ref. \cite{Yin98}.

We implement our Monte Carlo program by the BKL algorithm
\cite{BKL,Vot86} using a binary tree structure \cite{Blu95}.
In the algorithm, every trial leads to a jump.
At low temperatures, a large number of unsuccessful
trials inherent in the traditional Metropolis algorithm
can be avoided. This allows very long simulation times in our
system.

We first simulate adatom island diffusion with sizes $1 \le s \le 10^4$
at high temperature $T=1000$ K \cite{temperature}. 
Our data together with a fit
of $D$ from Ref. \cite{Kha96} (Eq. (36)) are shown in Fig. 1. 
For $s \gtrsim 10$ we clearly observe a crossover region where the 
effective scaling exponent behaves 
as predicted by Eq. (\ref{alpha}) (see the inset in Fig. 1).
For large islands, $\alpha$ finally approaches 
the limit $\alpha=1$ as predicted by theory \cite{Kha95,Kha96}. 
Due to the crossover, it is evident in Fig. 1 that for a limited
window of sizes, an effective exponent between $1 < \alpha < 3/2$
can be obtained.
Similar type of crossover region persists
at lower temperatures, and we find that
for example using the size window $100 \le s \le 1000$
we obtain values of $\alpha$ that only weakly depend on
temperature, {\it i.e.} 
$1.12 \le \alpha \le 1.23$ at $T=400, 500, 700$ and $1000$ K.
In particular, the overall
behavior of $D$ for large values of $s$ at $300$ K 
is in very good agreement with the behavior found in the experiments 
of Pai {\it et al.} \cite{Pai97} at room temperature
where $80 \le s \le 440$ (see Fig. 2)
($60 \le s \le 870$ for Ag).

The behavior of $D$ for smaller island sizes where
Eq. (\ref{alpha}) is not valid, is interesting.
There are clear size dependent oscillations present as also
reported by Fichthorn and Pal \cite{Fic98} in their simulations.
However, in the experiments such oscillations
are easily smeared out by size fluctuations \cite{Pai97}
as can be seen in Fig. 2 where the experimental data for $D$
follow closely the {\it average}
behavior of $D$ in the same regime.
At low temperatures there is a difference
for $D$ between small islands of sizes $n^2$ and $n^2\pm 1$,
where $n$ is an integer, in particular that 
$D(n^2)$ is much smaller than $D(n^2 + 1)$.
This is consistent with the notion that 
the square configurations are very stable and
therefore move slowly \cite{Vot86}.
However, for larger island this is oversimplified
since entropy must be taken into account.
In equilibrium, 
the probability for a given configuration to occur
$P(s,E) \propto \omega (s,E) e^{-\beta E}$, where $\omega (s,E)$
is the number of states of the island with size $s$ and energy $E$.
There is no degeneracy at $T=0$ for an $n^2$ configuration,
while the degeneracy of an ``excited state'' with one bond
broken ({\it e.g.} an adatom moving along the edge)
grows rapidly as a function of the island size.
Thus, for $n^2$ islands the contribution of the low-mobility
configuration to $D$ becomes less important when $n$ grows. 
Eventually, the oscillations dampen out and 
the continuum theory becomes valid \cite{arrhenius}.
At higher temperatures, this naturally occurs for smaller
islands already.

We now turn into discussion about the microscopic mechanisms
for island diffusion.
Most importantly, our simulation results 
indicate that there is a TD type of process involved
in the island motion in contrast to what was suggested
in Ref. \cite{Pai97} \cite{kinks}. 
This process in the present case is due to
{\it vacancy diffusion} inside the island.
This conclusion is supported by the observation that
the effective scaling exponent $\alpha$ approaches unity as the island
size increases even at room temperature
which indicates that the TD mode must be involved \cite{Kha95,Kha96}.
Moreover, we have explicitly checked the role of 
the PD and TD mechanisms at $1000$ K and $700$ K with $s \le 1000$.
We modified our model by first disallowing atoms to diffuse around corner
sites to prohibit the PD mechanism. In the second modification, we disallowed
the creation of vacancies in the island to prevent the TD mechanism from
operating.
Simulations of the two modified cases 
gave the scaling exponents $\alpha=1.02$ and $\alpha=1.48$,
in complete agreement with
the theoretical values for the
TD ($\alpha = 1$) and PD ($\alpha = 3/2$)
dominated island diffusion.

We have also measured the effective Arrhenius barriers
for island diffusion for $s=100,300,500,$ and $1000$,
and find that there is virtually no size dependence.
Interestingly enough, whether the PD or TD mechanism
is present makes also very little difference.
We have measured the barriers between $700$ K and $1000$ K
for the PD and the TD dominated cases with one of the
mechanisms suppressed as discussed in the section above, and
obtain $0.77$ eV and $0.79$ eV, respectively.
The Arrhenius barrier for the non-modified
case at $400$ K $\le T \le 1000$ K is $0.79$ eV. 
All these values are very close to 
the corresponding rate limiting process with 
$\Delta_{NN}=-2$.
This can be easily explained by microscopic considerations.
In the PD process, two bonds are broken when a particle goes from a kink
to a corner site \cite{Vot86}.
Symmetrically, in the TD process, the rate limiting step is the
creation of a vacancy where an atom having three neighbors
becomes a one-neighbor particle
{\it i.e.} a vacancy jumps into the island.
Therefore, jumps with $\Delta_{NN}=-2$
dominate the vacancy creation.

An interesting question for (100) metal surfaces concerns {\it
vacancy island} diffusion.
In our model, the energetics for vacancy islands is very
similar to the adatom case.
Symmetrically to adatom islands, vacancies are prevented to detach from a
vacancy island, but atoms can detach from the edge to the pit.
According to Eq. (\ref{energetics}) the barriers for vacancies
are then equivalent to the barriers for the adatoms,
except that the jumps inside the vacancy islands for
adatoms have $E_S=0.399$ eV
in contrast to $0.258$ eV for vacancies inside the islands.
However, this difference is not important in practice.
We have simulated vacancy island diffusion at various temperatures,
and the diffusion coefficients are the same as for the adatom 
islands within the statistical errors.
This is because 
the diffusion inside either an adatom or vacancy
island is {\it not} the rate limiting process.

To summarize, our model gives results in very good agreement with
experiments and theory 
and demonstrates that for at least Cu(100) surfaces,
vacancy diffusion within the islands contributes significantly
to the island mobility for larger islands
\cite{Vot86}. Another interesting feature not easily
seen in the experiments are the persistent oscillations in
$D$ at low temperatures that are due entropic reasons; in fact,
this is yet another example of the compensation effect
seen in many other systems. Our model predicts that vacancy island diffusion
on the Cu(100) surface is essentially similar to adatom
island diffusion since the rate-limiting mechanisms are
symmetric for both cases. 
Finally, we note that the present model is somewhat idealized in the
sense that the effect of other islands, surface steps etc. is neglected.
It would be of great interest to study these issues, as well as island
and vacancy diffusion on (100) surfaces of other metals to further clarify
the role of various microscopic mechanisms.\\

Acknowledgments: This work has been in part supported by the Academy
of Finland, and J.H. in part by the Finnish Graduate School
in Condensed Matter Physics.\\

$^\dagger$Corresponding author, e-mail address: 
{\tt jarkko.heinonen@helsinki.fi}.


\vskip1.0cm

FIG. 1. Adatom island diffusion coefficient $D$ {\it vs.} $s$
for $1 \le s \le 10^4$ at $T= 1000$ K ($a \approx 3.5$ {\AA}
is the lattice constant of copper). Stars
denote the results of our simulations, and the dashed line is a
fit to Eq. (36) from Ref. \cite{Kha96}
($R_{st}=5.0 \times 10^{-2}$ and $R_{su}=5.0 \times 10^{-4}$).
Error bars are of the size
of the symbols or smaller. The inset shows the effective
exponent $\alpha$ from Eq. (\ref{alpha}) using this fit.\\

FIG. 2. Adatom island diffusion coefficient $D$ for $1 \le s \le 10^4$
at $T=$ 1000, 700, 500, 400, and 300 K (from top to bottom).
For $T=$ 300 K, the $n^2$ configurations are shown with stars.
The dotted lines are just guides to the eye.
The dashed lines indicate fits to Eq. (36) of Ref. \cite{Kha96}
($R_{st}$ and $R_{su}$ are almost independent of $T$).
Error bars are of the size of the symbols or smaller except
at $T=$ 300 K for $s \gtrsim 100$ where the scatter in the data
indicates the errors.
The thick line at $T=300$ K shows the experimental results
of Ref. \cite{Pai97} for Cu. See text for details.\\


\begin{references}
\bibitem{temperature}
We note that the lattice-gas model used here is not expected to
be valid at high temperatures. However, for the present case
simulations of diffusion over a large temperature range
are important in order that the relevant low temperature
mechanisms be correctly identified (see also Ref. \cite{Fic98}).
\bibitem{arrhenius} From our data, we can estimate the limit $s_m$ 
where the oscillations become negligible by finding
the largest island
size for which $D(s-1)<D(s)$ still applies.
Using this definition we find in
temperature range 300 K $ \le T \le$ 1000 K that
$s_{m} \propto e^{\beta E_{m}}$, 
where the effective barrier $E_{m} \approx 0.26$ eV.
This is very close to the energy barrier needed to
excite an adatom from an $n^2$ configuration.
\end{references}
\end{document}